\documentclass[10pt, conference, compsocconf,fleqn]{IEEEtran} 

\usepackage{times}
\usepackage{alltt}
\usepackage{makecell}

  {\begin{quote}\footnotesize\begin{alltt}}%
  {\end{alltt}\end{quote}}

  {\begin{verse}\footnotesize}
  {\end{verse}}

\usepackage{amsmath}
\usepackage{amssymb,amsbsy,latexsym}
\usepackage{enumerate}
\usepackage{graphicx}
\usepackage{tabularx}
\usepackage{color, colortbl}
\usepackage[aboveskip=1cm]{subfig}

\begin{document}

\title{Analysis of buffering effects on hard real-time priority-preemptive wormhole networks}

\author{
\IEEEauthorblockN{Leandro Soares Indrusiak}
\IEEEauthorblockA{Department of Computer Science,\\
University of York, UK.\\
Email: leandro.indrusiak@york.ac.uk}\and

\IEEEauthorblockN{Alan Burns}
\IEEEauthorblockA{Department of Computer Science,\\
University of York, UK.\\
Email: alan.burns@york.ac.uk} \and

\IEEEauthorblockN{Borislav Nikoli{\'c}}
\IEEEauthorblockA{CISTER Research Centre, \\
ISEP/IPP, Porto, Portugal.\\
Email: borni@isep.ipp.pt} 
}

\maketitle

\begin{abstract}
There are several approaches to analyse the worst-case response times of sporadic packets transmitted over priority-preemptive wormhole networks. In this paper, we provide an overview of the different approaches, discuss their strengths and weaknesses, and propose an approach that captures all effects considered by previous approaches while providing tight yet safe upper bounds for packet response times. We specifically address the problems created by buffering and backpressure in wormhole networks, which amplifies the problem of indirect interference in a way that has not been considered by the early analysis approaches. Didactic examples and large-scale experiments with synthetically generated packet flow sets provide evidence of the strength of the proposed approach.

\end{abstract}

\section{Introduction}

Wormhole networks with priority-preemptive routers have been studied over more than two decades, mainly because of its small buffering overheads and of its potential time predictability. A number of approaches have attempted to calculate upper bounds to the latency of sporadic packets injected on such a network, but over the years each of them has been shown to be unsafe by increasingly complex counter-examples. The latest of them, presented by Xiong et al in~\cite{Xiong16}, shows that the approach by Shi and Burns~\cite{Shi08} is unsafe because it provides optimistic results under specific interference patterns (i.e. indirect interference caused by buffer backpressure). That development also shows that another recent approach must be unsafe as well, namely Kashif and Patel's in~\cite{Kashif2016}, as it claimed to be always tighter than (and therefore and upper-bounded by) Shi and Burns'. If it is upper-bounded by an optimistic approach, it must be optimistic too, and therefore unsafe.

Besides disproving Shi and Burns' analysis (and, indirectly, Kashif and Patel's),  Xiong et al proposed a new analysis to overcome the limitations of the preceding ones. To the best of our knowledge, their work is the current state-of-the-art for real-time analysis of priority-preemptive wormhole networks.

In this paper, we move the state-of-the-art one step further, showing that the work of Xiong et al~\cite{Xiong16} is unnecessarily pessimistic in its accounting of downstream indirect interference caused by buffer backpressure. And more importantly, we show with a counter-example that it is optimistic in its accounting of the overall indirect interference problem. We then propose a new analysis that is tighter in the accounting of indirect interference caused by backpressure, and safe on the overall accounting of indirect interference.  

The paper is organised as follows. Section \ref{WS} provides background on wormhole networks, followed by a formalisation of the problem of calculating upper-bounds of packet latencies over such networks in Section \ref{model}. A comprehensive survey on all relevant approaches to that problem is given in Section \ref{background}. The limitations of the state-of-the-art, and the proposed analysis that overcomes those limitations, are presented in Section \ref{sched1}. A quantitative and qualitative comparison showing the advantages of the proposed analysis is given in Section \ref{examples}, with selected examples as well as a large-scale evaluation with synthetic flowsets. The paper is closed by a brief overview of the changes to the state-of-the-art caused by the findings presented in this paper, as well as a preview of future developments.

\section{Wormhole Switching Networks}\label{WS}

Wormhole switching networks~\cite{Ni93} provide a good trade-off between time predictability and buffering overheads. Each packet in a wormhole network is divided into a number of fixed
size \emph{flits}, each of which is transmitted in parallel via a
number of wires that encode a single data item plus various flow
control signals.  The first flit of a packet (header flit) holds the packet size and the
routing information.  As the header advances along the specified route,
the remaining flits follow in a pipelined way. If the header flit
encounters a link already in use, it is blocked until the link becomes
available. In this situation, because network nodes have finite buffering capabilities, the second flit will be then blocked by the first one, and so on, until all flits stall in a process known as \emph{backpressure}. All flits of the packet will then remain buffered in the routers along the packet route until the header is released, so the pipelined transmission can continue. The smaller the buffers, the larger the number of routers that will store a given packet in a blockage scenario.

Since a packet can be stored by several routers and occupy multiple links at a time, the potential congestion over the network is increased. This makes it harder to predict the time it takes for a given packet to cross the network, because many of the links along its route may be blocked by other packets (e.g. as opposed to a store-and-forward network, where each packet uses only one link at a time). 

Wormhole networks are frequently used in Networks-on-Chip (NoCs)~\cite{Bjerregaard06}, because the possibility of small buffers is attractive due to limited overheads in silicon area and energy dissipation. In order to cope with the difficulties to predict packet latencies in wormhole NoCs, several arbitration mechanisms were proposed such as time-division multiplexing~\cite{Goossens05} and prioritised virtual channels (VCs)~\cite{Bolotin04,Shi08}. The first approach tries to avoid latency interference between packets by reserving link bandwidth to each packet flow. The second approach allows packets to interfere with each other but aims to quantify the upper bounds of that interference over each packet's latency. This paper follows the second approach, and uses a priority-preemptive NoC as a case study. Its findings, however, can be generalised to any wormhole switching network with priority-preemptive VCs.  
\begin{figure}[h]
  \centering
  \includegraphics*[width=\linewidth]{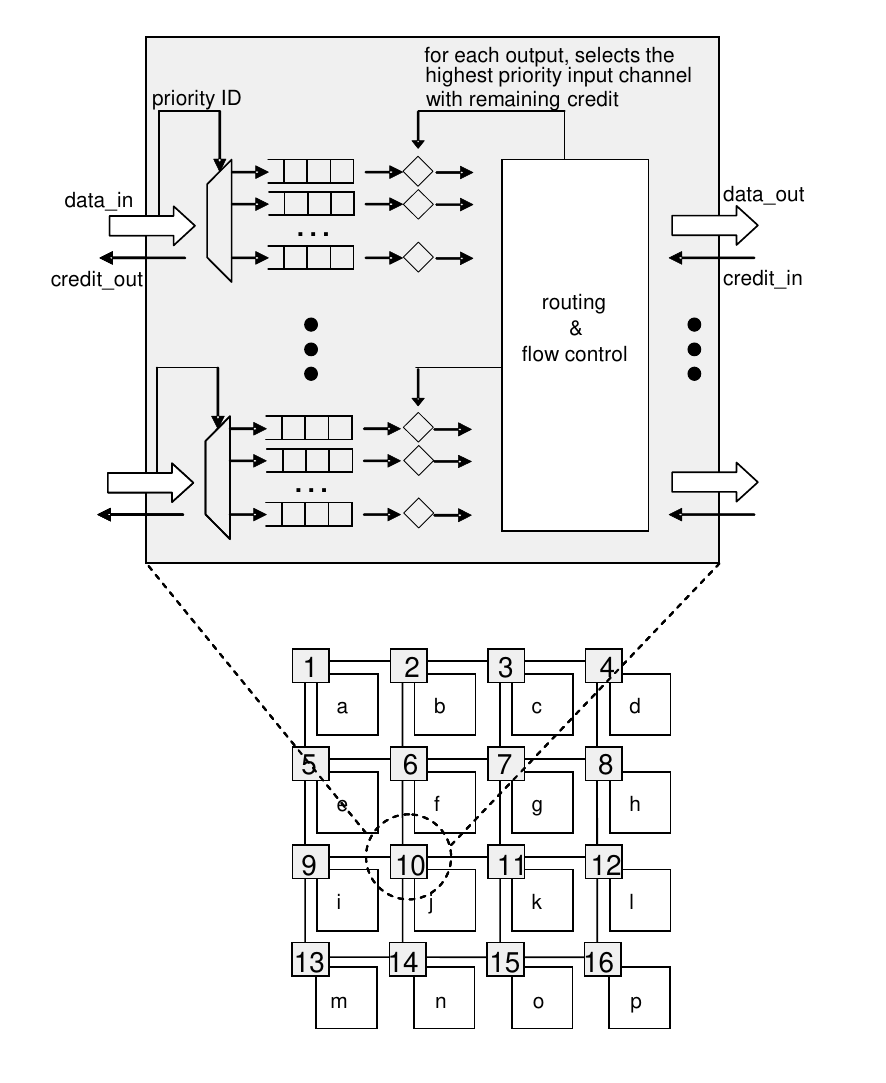}
  \caption{Wormhole on-chip network with 2D mesh topology and detail of a router with priority-driven virtual channels}
  \label{Figure:NoC}
\end{figure}

An implementation of a wormhole-based on-chip network is illustrated in Figure \ref{Figure:NoC}. Each core (labelled with lower-case characters) is connected to a network router through a network interface. Routers (labelled with numbers) are connected to each other following a mesh topology. This implementation follows the architectural templates presented in~\cite{Bolotin04}, where each router includes a flow controller based on priority preemptive VCs. By assigning priorities to packets, and by allowing high priority packets to preempt the transfer of low priority ones, network contention scenarios become more predictable and an upper bound to the packet latency can be found. The figure also shows the internal structure of the router of such a NoC. In each input port, a different FIFO buffer stores flits of packets arriving through different virtual channels (one for each priority level). The router assigns an output port for each incoming packet according to their destination. A credit-based approach~\cite{Bjerregaard06} guarantees that data is only forwarded from a router to the next when there is enough buffer space to hold it in the next hop. At any time, a flit of a given packet will be sent through its respective output port if it has the highest priority among the packets being sent out through that port, and if it has credits. If the highest priority packet cannot send data because it is blocked elsewhere in the network, the next highest priority packet can access the output link.

\section{Problem Description and System Model}\label{model}

The aim of this paper is, given a set of sporadic packet flows and their routes over a wormhole network with prioritised virtual channels, to find an upper bound to the latency of each of the packets. With such upper bound, we are able to test if a given system is schedulable: all packets can reach their destination while satisfying their timing requirements. To calculate such bounds, we require a more precise description of such a system.

Since we are interested in packet latencies, we take a communication-centric view of the system and focus on the traffic load imposed on the network. Thus, an application $\Gamma$ comprises $n$
real-time traffic-flows (or just \emph{flows} for short) $\Gamma$ =$
\{\tau_1,\tau_2, \ldots \tau_n\}$. Each flow $\tau_i$ gives rise to a
potentially unbounded sequence of \emph{packets}. A flow has a set
of properties and timing requirements which are characterised by a set
of attributes:
$\tau_i$ = ($P_i$, $C_i$, $T_i$, $D_i$, $J_i$, $\pi_i^s$,
$\pi_i^d$). All the flows which require timely delivery are either
periodic or sporadic. The lower bound interval on the time between
releases of successive packets is called the period ($T_i$) for the
flow. The maximum no-load network latency ($C_i$) is the maximum
duration of transmission latency when no flow contention exists, which is a function of the maximum number of flits $L_i$ of a packet of this flow, and the length of its route (defined below).

Each real-time flow also has a relative deadline ($D_i$) which is the upper bound restriction
on network latency, assumed to be $D_i \le T_i$.
Each flow also has a priority $P_i$; the
value 1 denotes the highest priority and larger integers denote lower
priorities. It also has a source and destination node on the network
($\pi_i^s$ and $\pi_i^d$).

Any flow can suffer a release jitter; $J_i$, which denotes the maximum
deviation of successive packet releases from the flow's period.
That is, a packet from $\tau_i$ will be released for transmission at
most  $J_i$ time units after its periodic tick, e.g. due to the time
it takes for its source node to execute the software task that generates it. 

We assume a wormhole-switching network with routers performing deterministic routing, credit-based flow control and priority-preemptive
arbitration of virtual channels implemented as multiple FIFO
input buffers (Figure \ref{Figure:NoC}). We assume that, for each transmission cycle, every output link of a router will transmit the flit of the highest-priority virtual channel that has flits to transmit to that link and that has credits.  

We model such a network as a set of nodes $\Pi = \{\pi_a,\pi_b, \ldots, \pi_z\}$, a set of
routers $\Xi = \{\xi_1,\xi_2, \ldots, \xi_m\}$, and a set of
unidirectional links $\Lambda = \{\lambda_{a1},\lambda_{1a}, \lambda_{12}, \lambda_{21}, \ldots, \lambda_{zm}, \lambda_{mz}\}$. 
The function $vc(\xi_i)$ denotes the number of virtual channels supported by router $\xi_i$, which in this model also means the number of priority levels it is able to distinguish. The function $buf(\xi_i)$ denotes the FIFO buffer size implementing a single virtual channel of that router (we assume all virtual channels to have the same buffer size, so the total buffer space per incoming link in a router $\xi_i$ is the product of $buf(\xi_i)$ and $vc(\xi_i)$). Assuming a homogeneous network, i.e. where all routers are equal, $buf(\Xi)$ denotes the virtual channel buffer size of any of them.

A network router is able to transmit flits over its links at a fixed rate. The amount of time taken by a router $\xi_i$ to transmit a flit over any of its links is represented by the link latency function $linkl(\xi_i)$. Assuming a synchronous and homogeneous network, the link latency of every link is $linkl(\Xi)$.

The route between any two nodes of the network is given by the function
$route(\pi_a,\pi_b) = \{\lambda_{a1},\lambda_{12},\ldots,\lambda_{mb}\}$, denoting the totally ordered subset of $\Lambda$ used to transfer
packets from node $\pi_a$ to node $\pi_b$. For convenience, we extend the notation of the
function $route$ to also represent the route of a packet from its source node to its destination:
$route(\tau_i) = route_i= route(\pi_i^s, \pi_i^d)$. The number of links of a route is given by $|route_i|$. Considering the routes of any two packets $\tau_i$ and $\tau_j$, we define a contention domain $cd_{i,j}$ as the ordered set of links shared by those packets: $cd_{i,j} = route_i \cap route_j$ . Finally, we define the funtion $order_{a, i}(\lambda_a, route_i)$ to denote the order of a link $\lambda_a$ over a route $route_i$ (i.e. 1 for first, 2 for second, etc.), and the respective convenience functions $first(route_i)$ and $last(route_i)$ to single out respectively the first and last link of the route of $\tau_i$.  

The goal of all the approaches reviewed in this paper, and of the one we propose, is to use (part of) the model presented above to calculate the worst case latency $R_i$ for each flow $\tau_i \in \Gamma$, which is an upper bound to the latencies of all packets produced by that flow. A system is then said to be schedulable if $R_i \le D_i$ for every $\tau_i \in \Gamma$.

\section{Related Work}\label{background}

The first approaches to upper-bound the latency of sporadic packets transmitted over wormhole networks with prioritised VCs were presented by Mutka~\cite{Mutka94} and Hary and Ozguner~\cite{Hary97}. Both approaches are based on fixed-priority schedulability analysis~\cite{Lehoczky89}. In~\cite{Hary97}, authors propose to consider the entire path of a given packet as a single shared resource, so that the worst case latency bound of a packet flow can be found by analysing the higher priority packet flows that share at least one link of its route. Kim et al~\cite{Kim98} noticed that neither of those approaches considered the effects of indirect interference, which happens when two packet flows do not share any network links but one of them can still have an impact on the latency bounds of the other (by affecting the behaviour of a third packet flow which shares links with both of them).

Lu et al presented the first approach to analyse worst case packet latencies in priority-preemptive wormhole NoCs in~\cite{Lu05}. Their analysis built on the notion of interference sets derived from the work of Kim et al in~\cite{Kim98}. The direct interference set $S_i^D$ of $\tau_i$ is the set of flows that have higher priority than $\tau_i$ and that share with it at least one network link (i.e. a non-empty contention domain):  $S_i^D = \{\tau_j \in \Gamma \mid  P_i < P_j ,  cd_{i,j} \ne \emptyset \}$. Similarly, the indirect interference set $S_i^I$ of $\tau_i$ is the set of flows that are not in $S_i^D$, but that interfere with at least one flow in that set (i.e. interfere with the flows that interfere with $\tau_i$, but not directly with $\tau_i$ itself):  $S_i^I = \{\tau_k \in \Gamma \mid  \tau_k \in  S_j^D, \tau_j \in  S_i^D, \tau_k \notin  S_i^D  \}$. Lu et al used the notion of interference sets to discriminate between packet flows that could not interfere with each other and would therefore be transmitted over the NoC in parallel. However, they incorrectly assumed that packets would experience their worst-case latency when they were released simultaneously.

In~\cite{Shi08}, Shi and Burns corrected that assumption and provided formulations for the upper-bound interference suffered by a given traffic flow $\tau_i$ considering both direct and indirect interferences. In the case of direct interference, they assume that a packet of $\tau_i$ may suffer interference from all packets of every flow $\tau_j \in S_i^D$ . The amount of interference on each ``hit" of a $\tau_j$ packet on $\tau_i$ is upper-bounded by $C_j$, and the number of ``hits" is bounded by the ratio surrounded by the ceiling function in Equation \ref{ShiBurnsDirInt} below. Adding up all interferences from all flows in $S_i^D$ leads to the total direct interference $I_i^D$ suffered by $\tau_i$:

\begin{equation}\label{ShiBurnsDirInt}
I_i^D \ = \sum_{\tau_j \in S_i^D} { \left\lceil
{\frac{R_i + J_j }{T_j}} \right\rceil C_j }
\end{equation}

Their analysis also considers the increased interference a packet from $\tau_i$ can suffer from two subsequent packets of a directly interfering flow $\tau_j$. This can happen if $\tau_j$ itself suffers interference from another flow, delaying the first of its packets to the point that it interferes on $\tau_i$ right before the second one causes interference (the so-called ``back-to-back hit"). They model that delay as an interference jitter $J_j^I$, which quantifies by how much, in the worst case, a packet of $\tau_j$ could be delayed by indirect interference. Clearly $J_j^I$ is upper-bounded by the actual interference suffered by $\tau_j$, so $J_j^I = R_j - C_j$. By adding the interference jitter $J_j^I$ to the regular release jitter $J_j$ in Equation \ref{ShiBurnsDirInt}, Shi and Burns formulated the composition of direct and indirect interference $I_i$ as:
 
\begin{equation}\label{ShiBurnsDirIndirInt}
I_i \ = \sum_{\tau_j \in S_i^D} { \left\lceil
{\frac{R_i + J_j  + J_j^I}{T_j}} \right\rceil C_j }
\end{equation}

Thus, according to Shi and Burns (SB) the worst case response time of a packet flow $\tau_i$ is:

\begin{equation}\label{SB}
R_i^{SB} \ = C_i + I_i = C_i + \sum_{\tau_j \in S_i^D} { \left\lceil
{\frac{R_i + J_j  + J_j^I}{T_j}} \right\rceil C_j }
\end{equation}

The authors do not make a clear statement about their assumptions regarding buffering and backpressure. Some of the subsequent work based on the SB analysis has assumed nodes with 2-flit buffers for each virtual channel, and a credit-based flow control to enable backpressure~\cite{ShiBurnsIndrusiak2010,IndrusiakReCoSoC2012}. Simulation-based experiments reported in those works provided evidence that the analysis was safe, albeit pessimistic at times.

An improvement to SB was proposed by Nikolic et al~\cite{NikolicArxiv2016}, which tried to reduce the pessimism resulting from the assumption that a packet occupies its complete route for the whole duration of its response time. The authors relied on the notion of a contention domain, representing the shared links between two packets, and analysed the interference patterns that can happen as a packet traverses links before, within and after each contention domain. With synthetically generated flowsets, they have shown an increased tightness in their improved analysis.      

In~\cite{Kashif2015}, Kashif et al proposed a similar improvement to SB, aiming to reduce its pessimism. They observed that the direct interference upper-bound of every ``hit" of a packet of $\tau_j$ on a packet of $\tau_i$ is often less than $C_j$, because packets do not necessarily interfere with each other over their complete routes. Their proposed SLA (stage-level analysis) calculated instead the interference on a link-by-link basis, resulting in tighter bounds for the worst case latency. The tightness of their bounds was evaluated using synthetically generated flow sets, showing improvements of up to 34\%. One limitation of that analysis, however, is the fact that it does not take into account the backpressure effects of a wormhole network. Instead, authors assumed that nodes always have enough buffering capacity for packets to flow (i.e. potentially infinite buffers). Despite this limitation, the authors claimed that their work superseded the SB analysis (as they assumed that SB also had the same limitation).

An improvement to SLA was presented by Kashif and Patel in~\cite{Kashif2016}, where they relaxed the assumption of infinite buffer capacity. They claim that their improved analysis will always be tighter and upper-bounded by SB. Experimental results show that their bounds are the same as SB with minimal buffer sizes, and get increasingly tighter in cases with larger buffer storage per VC.  

Xiong et al~\cite{Xiong16} have found a significant shortcoming in SB. They have identified using simulations that downstream indirect interference can sometimes cause a single packet of $\tau_j$ to directly interfere on $\tau_i$ by more than its basic latency $C_j$, disproving one of the SB assumptions. Specifically, they stated that a flit of a packet of $\tau_j$ may interfere multiple times on a packet of $\tau_i$ over multiple shared links (which we refer as the multiple interference problem), in case $\tau_j$ (1) suffers interference from a packet $\tau_k$ that does not interfere with $\tau_i$ and (2) shares links with $\tau_k$ downstream from the links it shares with $\tau_i$. 

To account for the downstream indirect interference problem, Xiong et al proposed a slightly different partitioning of indirect interference sets. They define the upstream indirect interference set $S_{I_i}^{{up}_j}$ as the set of flows $\tau_k \in  S_i^I$ that interfere with the flows $\tau_j \in S_i^D$ before $\tau_j$ interferes with $\tau_j$. Similarly, the downstream indirect interference set $S_{I_i}^{{down}_j}$ is the set of flows $\tau_k \in  S_i^I$ that interfere with the flows $\tau_j \in S_i^D$ after $\tau_j$ interferes with $\tau_j$. The notion of ``before" and ``after" used here refers to whether the contention domain between $\tau_k$ and $\tau_j$ (i.e. the links they share) appears upstream or downstream in $\tau_j$, in comparison with the contention domain between $\tau_i$ and $\tau_j$. For clarity, we rewrite the definition of those two sets using the notation introduced in Section \ref{model}:

\vspace{5mm}
$S_{I_i}^{{up}_j} = \{\tau_k \in S_i^I \cap S_j^D \mid order(first(cd_{jk}), route_j) < order(first(cd_{ij}), route_j) \}$
\vspace{5mm}

$S_{I_i}^{{down}_j} = \{\tau_k \in S_i^I \cap S_j^D \mid order(first(cd_{jk}), route_j) > order(first(cd_{ij}), route_j) \}$
\vspace{5mm}

Based on those two sets, Xiong et al defined two worst-case interference terms $ I_{ji}^{up}$ and $ I_{ji}^{down}$ to denote the worst case interference suffered by $\tau_j$ from flows $\tau_k$ that interfere with it, respectively, upstream or downstream from its contention domain with $\tau_i$:

\begin{equation}\label{XiongUp}
 I_{ji}^{up} \ =  \sum_{\tau_k \in S_{I_i}^{{up}_j}}{ I_{kj}}
\end{equation}

\begin{equation}\label{XiongDown}
 I_{ji}^{down} \ =  \sum_{\tau_k \in S_{I_i}^{{down}_j}}{ I_{kj}}
\end{equation}

They claimed that the indirect interference jitter $J_j^I$ defined in the SB analysis is only caused by upstream indirect interference, so they redefine it as $J_j^I = I_{ji}^{up}$. Then, they claim that the downstream indirect interference suffered from every $\tau_j$ manifests itself as direct interference over $\tau_i$, so they add $ I_{ji}^{down}$  to $C_j$ in their proposed worst case response time of a packet flow $\tau_i$, which we refer as XLWX: 

\begin{equation}\label{Xiong}
R_i^{XLWX}  = C_i  + \sum_{\tau_j \in S_i^D} { \left\lceil
{\frac{R_i + J_j  + I_{ji}^{up}}{T_j}} \right\rceil 
( C_j + I_{ji}^{down} ) }
\end{equation}

\vspace{3mm}

\section{Proposed Analysis}\label{sched1}

The key motivation for the approach presented in this paper is the treatment of the indirect interference problem in the XLWX analysis. While the Xiong et al have clearly identified a type of interference that has not been considered in the previous approaches, we argue that their analysis approach does not properly address the indirect interference effects that happen in wormhole networks, and does not provide a safe and tight upper bound to packet latency. 

Firstly, their handling of downstream indirect interference as if it were direct interference is unnecessarily pessimistic, so we aim to provide a tighter analysis by considering more carefully the impact of the multiple interference problem. Secondly, we disagree with their claim that the indirect interference jitter $J_j^I$ defined in the SB analysis is only caused by upstream indirect interference, and we show that their approach of considering only the upstream indirect interference set to calculate that jitter term is unsafe.

Let us carefully revisit the multiple interference problem, caused by the downstream indirect interference identified in~\cite{Xiong16}: a single packet of $\tau_j$ can directly interfere on $\tau_i$ by more than its basic latency $C_j$ when it suffers interference from any packet $\tau_k$ that does not interfere with $\tau_i$ and shares links with $\tau_k$ downstream from the links it shares with $\tau_i$. In this situation, every time $\tau_j$ is blocked by $\tau_k$, it can allow $\tau_i$ to flow through the network and potentially overtake $\tau_j$ flits that had already blocked it earlier. This effect is depicted in Figures 2 and 3 of~\cite{Xiong16}. XLWX analysis correctly takes into account that the amount of additional interference that $\tau_i$ can suffer from $\tau_j$ is upper-bounded by the amount of time that $\tau_i$ is allowed to overtake $\tau_j$ (and subject itself to additional interference), which is in turn upper-bounded by the downstream indirect interference that $\tau_j$ can suffer from any $\tau_k$ (which is expressed by  $I_{ji}^{down}$, as shown in Equation \ref{XiongDown}). In a simple example with five flows, Xiong et al show in~\cite{Xiong16} that the SB analysis does not capture the multiple interference problem caused by downstream indirect interference, and thus produces an optimistic result, while XLWX analysis provides an upper bound in all cases.

\begin{figure}[!htb]
  \centering
  \includegraphics*[scale=1.0]{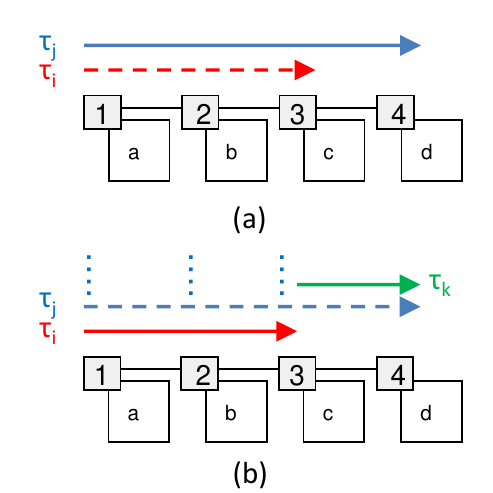}
  \caption{Downstream indirect interference}
  \label{Figure:Interference}
\end{figure}

We will focus on the example presented by~\cite{Xiong16} in the next section. But first, we use a smaller and more didactic example with only three flows $\tau_i$, $\tau_j$ and $\tau_k$, as shown in Figure \ref{Figure:Interference}. Assume that $\tau_i$ and $\tau_j$ have much larger periods and longer packets (therefore larger $C$) than $\tau_k$, and that $\tau_k$'s releases are not in phase with the other two. The priority order has $\tau_i$ with the lowest and $\tau_k$ with the highest priority. In Figure \ref{Figure:Interference}(a), $\tau_i$ and $\tau_j$ are released at the same time from core \emph{a}, and the higher priority $\tau_j$ gains access to the network, blocking $\tau_i$. 

In Figure \ref{Figure:Interference}(b), a packet of $\tau_k$ is then released and interferes with $\tau_j$ (downstream from its contention domain $c_{ij}$ with $\tau_i$). Since $\tau_k$ has the highest priority, it stops $\tau_j$'s flits from using the link between routers \emph{3} and \emph{4}, which generate backpressure on all subsequent flits of that packet of $\tau_j$, forcing them to stay buffered along the route (depicted as stacked square dots) all the way to the source in core \emph{a}. Once $\tau_j$ flits stop using the links on $\tau_i$'s route,  $\tau_i$ then becomes the highest priority flow with buffer credits so the routers starts transmitting its flits.

When $\tau_k$ finishes, the scenario returns to the situation depicted in Figure \ref{Figure:Interference}(a), where only $\tau_j$ flows through the network. However, before new flits of $\tau_j$ can flow out of core \emph{a}, its buffered flits must first make way and release the backpressure along the route. This is key to the downstream interference problem: it is those buffered flits of $\tau_j$, which have already caused interference on $\tau_i$ when they were first released out of core \emph{a}, that will again cause interference and as a consequence will delay $\tau_i$ by more than $\tau_j$'s no-load latency $C_j$. From now onwards, we refer to this effect as \emph{buffered interference}. 

By understanding the notion of buffered interference, one can clearly see that the intuition behind XLWX analysis holds: the interference beyond $C_j$ imposed by $\tau_j$ on $\tau_i$ will never be larger than the amount of downstream interference that $\tau_j$ suffers from $\tau_k$, since that is the maximum amount of interference from $\tau_j$ that could be buffered along its way. Thus, by adding the maximum downstream interference $I_{ji}^{down}$ to $C_j$ Xiong et al effectively provides a safe upper-bound to the multiple times $\tau_j$ can interfere with $\tau_i$.

We claim, however, that such upper bound is unnecessarily pessimistic, given that the amount of buffered interference will also be upper-bounded by the maximum amount of buffer space along the route of $\tau_j$. Furthermore, we claim that the amount of buffered interference of a single packet of $\tau_j$ that can interfere multiple times with $\tau_i$ is proportional to the length of their contention domain $cd_{ij}$. The intuition behind our claims is based on two observations regarding the behaviour of a $\tau_j$ packet once it starts flowing again after the end of a downstream interference ``hit" by $\tau_k$:

\begin{itemize}
  \item the flits of $\tau_j$ stored in buffers of routers that are upstream to the contention domain $cd_{ij}$ have not yet caused any interference on $\tau_i$, so they won't contribute to the multiple interference problem unless there are more downstream interference ``hits" during the lifetime of the packet.

\item the flits of $\tau_j$ stored in buffers of routers that are downstream to the contention domain $cd_{ij}$ will not cause any further interference on $\tau_i$, so they will not contribute to the multiple interference problem.
\end{itemize}

This shows that, for each downstream interference ``hit", the only $\tau_j$ flits that can interfere more than once on $\tau_i$ are those stored in the buffers along their contention domain $cd_{ij}$. Based on that, we can define a formulation for the maximum buffered interference over the contention domain $cd_{ij}$:

\begin{equation}\label{IndrusiakBufferedInterference}
bi_{ij} \ = buf(\Xi) . linkl(\Xi) . | cd_{ij} |
\end{equation}

and a new upper-bound for the downstream indirect interference:

\begin{equation}\label{IndrusiakDown}
 I_{ji}^{down} \ =  \sum_{\tau_k \in S_{I_i}^{{down}_j}}{ 
\left\lceil
{\frac{R_j + J_k }{T_k}} \right\rceil bi_{ij}
}
\end{equation}

The ceiling function in Equation \ref{IndrusiakDown} determines the number of hits suffered by $\tau_j$ from every $\tau_k$ in the downstream indirect interference set of $\tau_i$, which is  multiplied by the buffered interference of each hit calculated by Equation \ref{IndrusiakBufferedInterference}, i.e. the time it takes for the flits of $\tau_j$ buffered along $cd_{ij}$ to flow and potentially hit $\tau_i$ again. That time is given by the product of the amount of buffer space per router on the virtual channel of $\tau_j$ ($buf(\Xi)$), the time it takes for each one of the buffered flits to cross a network link ($linkl(\Xi)$) and the number of links in the contention domain of $\tau_j$ and $\tau_i$ ($| cd_{ij} |$). 

While the proposed upper bound in Equation \ref{IndrusiakDown} is often tighter than the one presented in ~\cite{Xiong16}, that is not always the case. In the cases that the downstream interference on $\tau_j$ is not large enough to generate backpressure to fill up all the buffers along the contention domain $cd_{ij}$, it is likely that the maximum buffered interference $bi_{ij}$ could be larger than the maximum downstream interference $C_k$, making the original analysis tighter. Therefore, we rewrite Equation \ref{IndrusiakDown} to use, for every downstream interference hit, the smallest value between $bi_{ij}$ and $C_k$: 

\begin{equation}\label{IndrusiakDownImproved}
 I_{ji}^{down} \ =  \sum_{\tau_k \in S_{I_i}^{{down}_j}}{ 
\left\lceil
{\frac{R_j + J_k }{T_k}} \right\rceil  min(bi_{ij}, C_k)
}
\end{equation}

As we will show in the next section, the claim from~\cite{Xiong16} that the indirect interference jitter $J_j^I$ defined in the SB analysis is only caused by upstream indirect interference is wrong. So, we argue that it is not safe to use the interference jitter term defined in Equation \ref{XiongUp}. As shown in~\cite{Shi08}, both upstream and downstream indirect interference can cause two successive packets of $\tau_j$ to arrive closer than expected to each other (i.e. back-to-back hit). Following the reasoning in the SB analysis, we assume $J_j^I = R_j - C_j$. Thus, using $I_{ji}^{down}$ as defined in Equation \ref{IndrusiakDownImproved}, we have the upper-bound latency according to the proposed analysis (referred as IBN) given by:

\begin{equation}\label{IBN}
R_i^{IBN} \ = C_i + \sum_{\tau_j \in S_i^D} { \left\lceil
{\frac{R_i + J_j  + J_j^I}{T_j}} \right\rceil ( C_j +  I_{ji}^{down}) }
\end{equation}

\vspace{3mm}

\section{Comparative Analysis}\label{examples}

In this section, we first use three selected examples to compare the proposed analysis to the XLWX and SB analyses, aiming to show it is still safe in counter-examples that provide evidence via simulation that the baselines are optimistic. We also show that the proposed analysis is tighter than XLWX in a typical downstream indirect interference example. For the sake of simplicity, the figures describing the examples depict small networks with only a few nodes, but simulations were performed using a cycle-accurate full NoC simulator configured for single-cycle header routing and single-cycle link latency (including a link connecting a core to its respective router).  

We then provide additional evidence on the tightness of the proposed analysis by performing a large-scale comparison using synthetically-generated flowsets of increasing load.

\subsection{Example 1}\label{example1}

\begin{figure}[h]
  \centering
  \includegraphics*[scale=1.0]{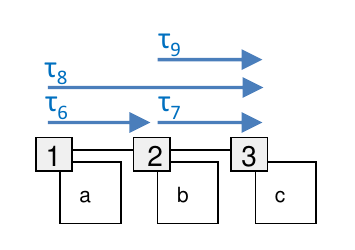}
  \caption{Flow routes for Example 1}
  \label{Fig:NikolicExample}
\end{figure}

To highlight the problems on the formulation of the indirect interference jitter in~\cite{Xiong16}, we introduce a small example with only four packet flows, referred as Example 1.  Table \ref{Table:NikolicExample} shows the parameters of each flow and Figure \ref{Fig:NikolicExample} shows their routes.  We then applied three analyses approaches to this example (SB, XLMX and IBN), which produced latency upper-bounds $R$ for each flow, as shown in Table \ref{Table:NikolicExampleResults}.

\begin{table}[ht]
\centering
\caption{Flow parameters for Example 1}
\label{Table:NikolicExample}
\begin{tabular}{llllll}
\hline
\\
flow & C (L, $\mid route \mid$)  & T   & D   & J & P \\
\\
\hline
\hline
\\
$\tau_6$    & 14 (12, 3)  & 1000 & 1000 & 0 & 1 \\
$\tau_7$    & 52  (50, 3) & 208 & 208 & 0 & 2 \\
$\tau_8$    & 103 (100 ,4) & 257 & 257 & 0 & 3 \\
$\tau_9$    & 52 (50, 3) & 1000 & 250 & 0 & 4 \\
\hline

\end{tabular}
\end{table}

In this example, the proposed analysis produces the same results as SB. This is expected, since it does not have to take into account the multiple interference problem because there is no downstream indirect interference. Both SB and the proposed analysis provide an upper-bound to the highest values found using simulation. XLWX, however, produces an optimistic value for the latency upper-bound of $\tau_9$. One of the simulation scenarios that produces a latency higher than the XLWX upper-bound for $\tau_9$ is as follows:  $\tau_7$ and $\tau_8$ are released at t=0, $\tau_6$ is released at t=50 and $\tau_9$ released at t=61. 

The issue with XLWX is their upstream indirect interference jitter $I_{ji}^{up}$, which is unable to properly capture all the indirect interference effects. XLWX analysis could be corrected by using $J_j^I = R_j - C_j$ instead of $I_{ji}^{up}$, which would make it safe but never tighter than the proposed analysis. 

\begin{table}[h]
\centering
\caption{Analysis and simulation results for Example 1}
\label{Table:NikolicExampleResults}
\begin{tabular}{lllll}
\hline
\\
flow & $R^{SB}$  & $R^{XLWX}$ & $R^{IBN}$ & $R^{sim}$  \\
\\
\hline
\hline
\\
$\tau_6$    & 14   & 14   & 14          & 14                  \\
$\tau_7$   & 52  & 52   & 52                & 52           \\
$\tau_8$    & 169 & 169  & 169        & 153          \\
$\tau_9$    & 362 & 207  & 362             & 302         \\ 
\hline

\end{tabular}
\end{table}

\subsection{Example 2}\label{example2}

\begin{figure}[!htb]
  \centering
  \includegraphics*[scale=1.0]{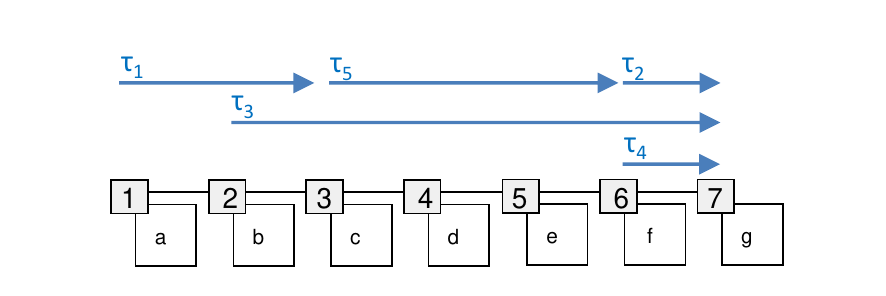}
  \caption{Flow routes for Example 2~\cite{Xiong16} }
  \label{Fig:XiongExample}
\end{figure}

Now, we revisit the the example with five traffic flows presented by~\cite{Xiong16}, and refer to it as Example 2. Table \ref{Table:XiongExample} shows the parameters of each flow, while Figure \ref{Fig:XiongExample} shows their routes.

\begin{table}[ht]
\centering
\caption{Flow parameters for Example 2~\cite{Xiong16}}
\label{Table:XiongExample}
\begin{tabular}{llllll}
\hline
\\
flow & C (L, $\mid route \mid$)  & T   & D   & J & P \\
\\
\hline
\hline
\\
$\tau_1$    & 30 (27, 4)  & 150 & 100 & 0 & 1 \\
$\tau_2$    & 30  (28, 3) & 150 & 100 & 0 & 2 \\
$\tau_3$    & 150 (144, 7) & 400 & 300 & 0 & 3 \\
$\tau_4$    & 100 (98, 3) & 600 & 550 & 0 & 4 \\
$\tau_5$    & 100 (96,5) & 300 & 250 & 0 & 5 \\
\hline
\end{tabular}
\end{table}

We again applied all three analyses approaches to this example (SB, XLMX and IBN), which produced latency upper-bounds $R$ for each flow, as shown in Table \ref{Table:XiongExampleResults}. To provide evidence that the proposed analysis can capture the influence of the buffer and contention domain sizes on the downstream indirect interference, we tabulate the results of the proposed analysis considering different buffer sizes (2 and 10-flit buffers per VC), which are identified by the subscript $b = buffer size$. We also produced simulation results for the same buffer sizes used for the proposed analysis, and tabulated the worst observed latency for each flow (using the same subscripts to identify the buffer sizes used in each simulation scenario).

\begin{table}[h]
\centering
\caption{Analysis and simulation results for Example 2}
\label{Table:XiongExampleResults}
\begin{tabular}{lllllll}
\hline
\\
flow & $R^{SB}$  & $R^{XLWX}$ & $R^{IBN}_{b=10}$ & $R^{IBN}_{b=2}$ & $R^{sim}_{b=10}$& $R^{sim}_{b=2}$  \\
\\
\hline
\hline
\\

$\tau_1$    & 30  & 30   & 30          & 30         & 30            & 30           \\
$\tau_2$   & 30  & 30   & 30          & 30         & 30            & 30           \\
$\tau_3$    & 270 & 270  & 270         & 270        & 233           & 205          \\
$\tau_4$    & 520 & 340  & 520         & 520        & 300           & 300          \\
$\tau_5$    & 250 & 310  & 520         & 262        & 264           & 247         \\
\hline
\end{tabular}
\end{table}

Xiong et al~\cite{Xiong16} used the example above to show that SB analysis is unsafe, since it produces optimistic results for $\tau_5$ in the large buffer simulation scenario, while their analysis is safe for all flows. Table \ref{Table:XiongExampleResults} shows that the proposed analysis is also safe for all flows, as its upper bounds are the same or above the results found using simulation. It also shows that the proposed approach is tighter than XLWX for the 2-flit buffer scenario, but it is less tight in the case of 10-flit buffers. This is because XLWX overestimates the downstream indirect interference but underestimates the overall indirect interference (as shown in the previous example), and in this case the underestimation is such that makes XLWX analysis seem tighter.

\subsection{Example 3}\label{example3}

We now introduce a third example to better show the tightness of the proposed analysis when compared to XLWX. In this example, we use only three flows and reuse the routes of $\tau_2$, $\tau_3$ and $\tau_5$ from Example 2, so that the scenario is the same as the one shown in Figure \ref{Fig:XiongExample}, but discarding $\tau_1$ and $\tau_4$. Such changes remove the possibility of upstream indirect interference, therefore a fair comparison between both analyses can be done (i.e. by removing the possibility of an unsafe result in XLWX). We then use the flow parameters from Table \ref{Table:IndrusiakExample} to highlight the effects of the downstream indirect interference of $\tau_2$ over $\tau_5$ through $\tau_3$.

\begin{table}[ht]
\centering
\caption{Flow parameters for Example 3}
\label{Table:IndrusiakExample}
\begin{tabular}{llllll}
\hline
\\
flow & C (L, $\mid route \mid$)  & T   & D   & J & P \\
\\
\hline
\hline
\\
$\tau_2$    & 62  (60, 3) & 200 & 200 & 0 & 1 \\
$\tau_3$    & 204 (198, 7) & 4000 & 4000 & 0 & 2 \\
$\tau_5$    & 132 (128,5) & 6000 & 6000 & 0 & 3 \\
\hline

\end{tabular}
\end{table}

The results in Table \ref{Table:IndrusiakExampleResults} show, as expected, that both the proposed analysis and XLWX provide upper-bounds to the values found using simulation while SB provides optimistic bounds. This time, however, the proposed analysis has much tighter results than XLWX for $\tau_5$ (348 vs 460 for 2-flit buffer networks, or 396 vs 460 for 10-flit buffer networks). This happens because in this example the excessive pessimism introduced by XLWX in its accounting of the multiple interference problem is not offset by its incorrect optimism on the accounting of indirect interference. 

The results for the proposed analysis using different buffer sizes show that the common practice of using small buffers in wormhole networks is also advantageous in terms of time predictability, since smaller buffers allow the proposed analysis to have tighter bounds because of the limited amount of buffered interference that can build up in the network.

\begin{table}[h]
\centering
\caption{Analysis and simulation results for Example 3}
\label{Table:IndrusiakExampleResults}
\begin{tabular}{lllllll}
\hline
\\
flow & $R^{SB}$  & $R^{XLWX}$ & $R^{IBN}_{b=10}$ & $R^{IBN}_{b=2}$ & $R^{sim}_{b=10}$& $R^{sim}_{b=2}$  \\
\\
\hline
\hline
\\

$\tau_2$   & 62  & 62   & 62          & 62         & 62            & 62           \\
$\tau_3$    & 328 & 328  & 328         & 328        & 324           & 324          \\
$\tau_5$    & 336 & 460  & 396         & 348        & 352           & 336         \\
\hline
\end{tabular}
\end{table}

\subsection{Large-Scale Evaluation}\label{eval}

\begin{figure*}[ht]
  \centering
  \includegraphics*[scale=1.0]{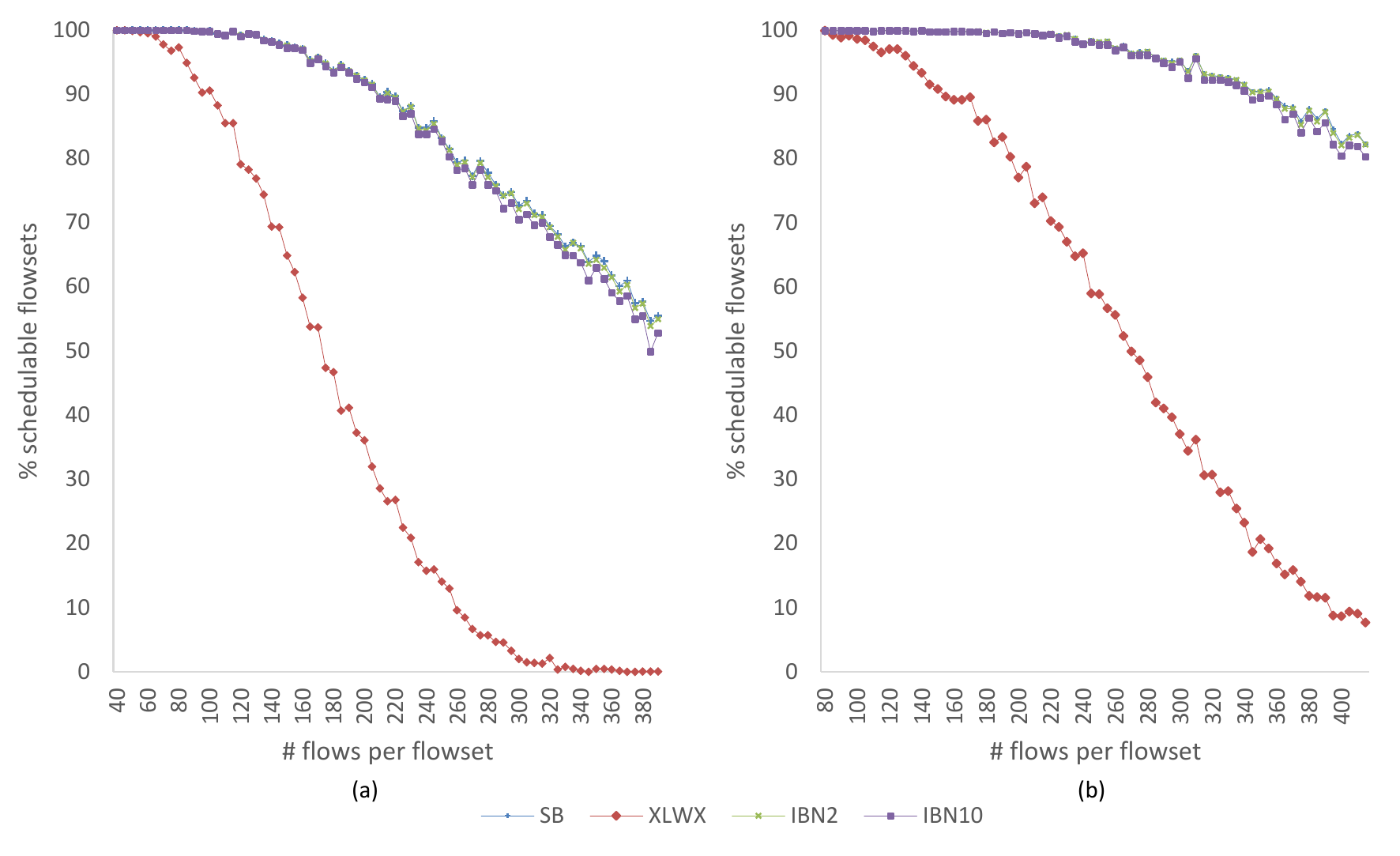}
  \caption{Schedulability results for the proposed analysis and the SB and XLXB baselines, for (a) 4x4 NoC and (b) 8x8 NoC. Each point represents 1000 flowsets, each of them with the number of flows indicated over the Y-axis.}
  \label{Fig:SchedPlots}
\end{figure*}

\begin{figure*}[ht]
  \centering
  \includegraphics*[scale=0.95]{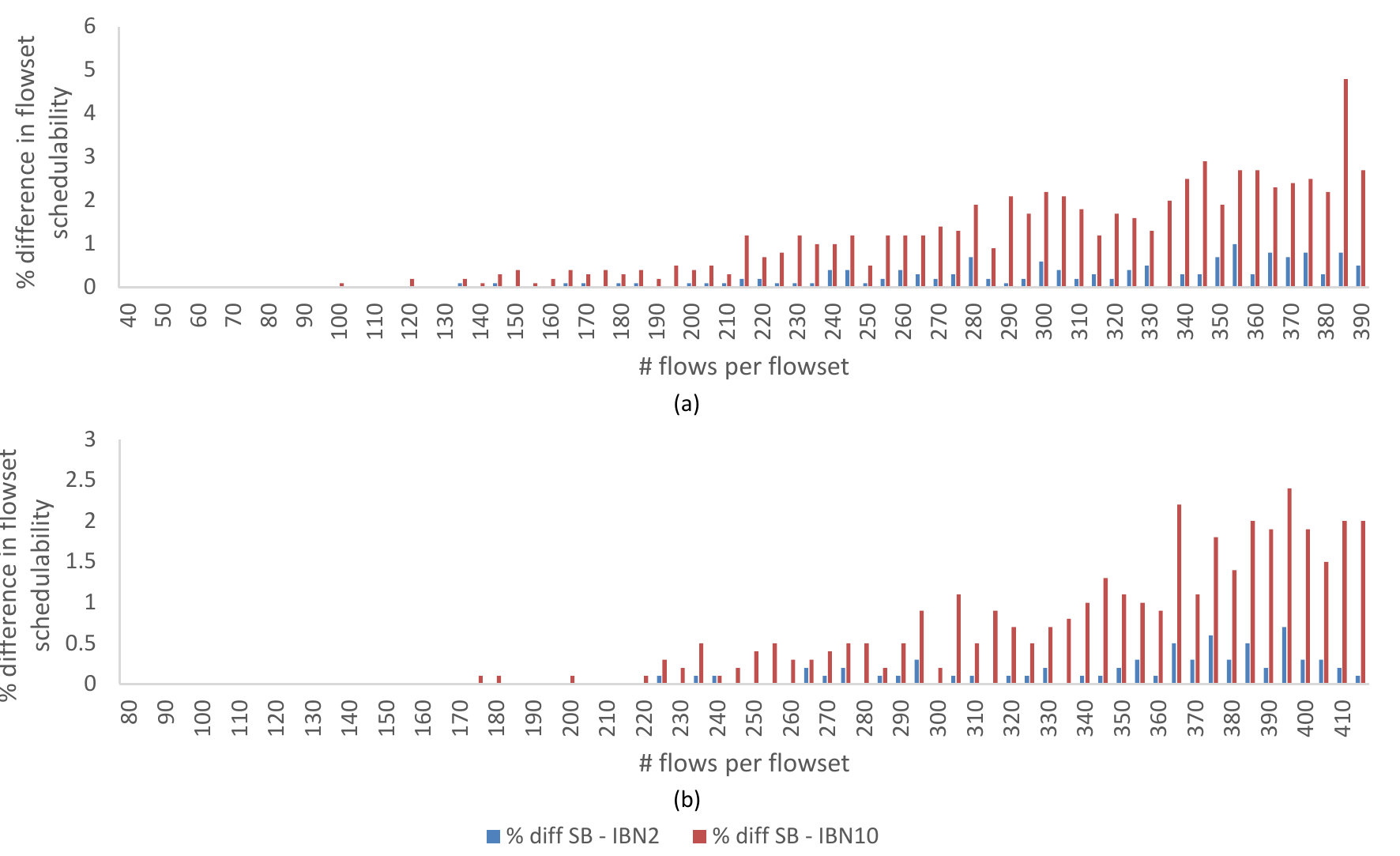}
  \caption{Difference in schedulability between SB and the proposed analysis considering different buffer sizes, 2-flit and 10-flit buffers, for (a) 4x4 NoC and (b) 8x8 NoC. }
  \label{Fig:Diff}
\end{figure*}

We now perform large-scale comparisons using synthetically-generated sets of packet flows, and two priority-preemptive wormhole network-on-chip platforms, a 16-core (4x4) and a 64-core (8x8). Both platforms follow the template shown in Figure \ref{Figure:NoC} with 2D-mesh topology, as well as deterministic XY routing, single-cycle routing capabilities, and operating frequency of 100 MHz. 

We use flowsets of increasing workload by varying the number of flows in each set. Each set, however, is composed by flows with the similar characteristics: periods uniformly distributed between 0.5 s and 0.5 ms, and maximum packet lengths uniformly distributed between 128 and 4096 flits. Sources and destinations of packet flows are randomly selected, so the average route is longer in the larger platform. Rate-monotonic priority assignment is used despite sub-optimality, given that no optimal assignment is known for this problem.

On each of the plots in Figure \ref{Fig:SchedPlots}, we show over the X-axis the number of flows in each generated flowset, and the Y-axis the percentage of flowsets found schedulable out of a set of 1000 flowsets. We have plotted, respectively for the 4x4 in Figure \ref{Fig:SchedPlots}(a) and 8x8 NoC in Figure \ref{Fig:SchedPlots}(b), the percentage found by SB and XLWX analyses, as well as the proposed analysis considering network routers with 2-flit buffers per VC (referred to as IBN2) and with 10-flit buffers per VC (referred as IBN10). 

In both plots it can be seen that the difference between XLWX and the other analysis becomes extremely large as the workload on the network increases. This is due to the large amount of pessimism that results from the treatment of indirect interference as if it were direct interference. As expected, the proposed analysis tightly follows the SB analysis, and in most of cases their lines appear indistinguishable on the plot. Looking more closely, IBN2 and IBN10 are always slightly more conservative than SB, and this increases with the increase of the workload. Figure \ref{Fig:Diff} shows that the difference is never more than a few percent points. This hints that the downstream indirect interference patterns are not easily found, which explains why their effects have not been seen in previous simulation-based evaluation of SB in~\cite{ShiBurnsIndrusiak2010,IndrusiakReCoSoC2012}. It also clearly shows that the difference is larger in the cases with 10-flit buffers, corroborating the statement made in the previous subsections that large buffers decrease the predictability of the network.

\section{Conclusions}\label{con}

This paper has reviewed the state-of-the-art in real-time analyses of priority-preemptive wormhole networks, and has proposed a novel analysis that extends the state-of-the-art by carefully modelling the effects of buffering on indirect interference. The work aimed to capture the multiple interference problem caused by downstream indirect interference, identified in~\cite{Xiong16}. We have improved over the analysis of~\cite{Xiong16} by showing the relation between the impact of the multiple interference problem and the amount of buffer space over the network links shared by the interfering packet flows. We have also shown with a counter-example that their treatment of indirect interference is incorrect, which makes the proposed analysis the only known approach that is safe for all known buffering and interference effects in priority-preemptive wormhole networks. Table \ref{Table:Comparison} further enforces this point by presenting a comparative overview of the existing analyses and their coverage of the buffering and interference effects:

\begin{table*}[ht]
\caption{Comparative overview of real-time analyses for priority-preemptive wormhole networks}
\centering
\label{Table:Comparison}
\begin{tabular}{l  c c  c  c   c  c c}
\hline
\\
analysis & \makecell{direct \\ interference}  & \makecell{indirect \\ interference} & backpressure &  \makecell{non-zero \\ critical \\ instant} &  \makecell{sub-route \\ interference} &  \makecell{downstream \\ mutiple \\ interference} & safe \\
\\
\hline
\hline
\\
Mutka~\cite{Mutka94} & Y & N & Y & N & N & N & N \\
\\
Hary and Ozguner~\cite{Hary97} & Y & N & Y & N & N & N & N \\
\\
Kim et al~\cite{Kim98} & Y & Y & Y & N & N & N & N \\
\\
Lu et al~\cite{Lu05} & Y & Y & Y & N & N & N & N \\
\\
Shi and Burns~\cite{Shi08} & Y & Y & Y & Y & N & N & N \\
\\
Nikolic et al~\cite{NikolicArxiv2016} & Y & Y & Y & Y & Y & N & N \\
\\
Kashif et al~\cite{Kashif2015}  & Y & Y & N & Y & Y & N & Y \\
\\
Kashif and Patel~\cite{Kashif2016}  & Y & Y & Y & Y & Y & N & N \\
\\
Xiong et al~\cite{Xiong16}  & Y & Y & Y & Y & N & Y & N \\
\\
Indrusiak et al (proposed)  & Y & Y & Y & Y & N & Y & Y \\
\\
\hline
\end{tabular}
\end{table*}

\begin{itemize}
  \item \emph{direct interference}: whether the analysis takes into account direct interference
  \item \emph{indirect interference}: whether the analysis takes into account indirect interference that reduces the time between subsequent packets of a given flow
  \item \emph{backpressure}: whether the analysis considers finite buffers, and the backpressure effect that stops flits from moving when buffers are full
  \item \emph{non-zero critical instant}: whether the analysis takes into account that there are scenarios where the critical instant is not when all flows are released simultaneously
  \item \emph{sub-route interference}: whether the analysis can handle an interference granularity that is smaller than the complete route of the packet (does not affect the safety of the analysis, only its tightness)
  \item \emph{downstream multiple interference}: whether the analysis captures the multiple interferences that can be caused by downstream indirect interference
  \item \emph{safe}: whether the analysis is safe, i.e. there are no known counter-examples showing that it produces optimistic results 
\end{itemize}

The results we presented have also provided additional evidence of the advantages of using small buffer sizes in wormhole networks, since tighter analysis bounds can be obtained for networks with smaller buffers. In other words, while they may provide improvements on average-case performance, larger buffers can only increase the worst-case performance of wormhole networks.

There are ways to make the proposed analysis tighter. For instance, it is possible to use the number of busy periods at the priority level of the flow causing indirect interference, rather than the number of hits it takes from downstream interfering flows, to calculate $I^{down}_{ji}$ in Equation \ref{IndrusiakDown}. It is also possible to reduce $C_k$ in Equation \ref{IndrusiakDownImproved} by the amount of time it takes for the backpressure to go from the first link of $cd_{jk}$ and the last link of $cd_{ij}$. Such approaches will make the analysis more complex, and are therefore left for future work. Additional work is currently ongoing to improve the tightness of the proposed analysis by supporting sub-route granularity (following~\cite{NikolicArxiv2016}), and to investigate the possibility of handling downstream indirect interference within a mixed-criticality approach such as~\cite{BurnsRTSS2014,IndrusiakECRTS2015}.

\subsection*{Acknowledgements}
The research described in this paper is funded, in part, by EPSRC-funded MCC project (EP/K011626/1). No new primary data were created during this study.

\bibliographystyle{plain}
\bibliography{refs}

\end{document}